\documentclass{aa}
\usepackage[comma,authoryear]{natbib}
\usepackage{graphics}
\usepackage[latin1]{inputenc} % to insert accentuated characters

\begin{document}

   \title{On the correlation between Ca and H$\alpha$ solar emission and consequences for stellar activity observations}
   \titlerunning{Correlation between Ca and H$\alpha$ solar emission and stellar activity}

   \author{N. Meunier \inst{1}, X. Delfosse \inst{1} 
	  }
   \authorrunning{Meunier and Delfosse}

   \institute{
Laboratoire d'Astrophysique de Grenoble, Observatoire de Grenoble, Universit\'e Joseph Fourier, CNRS, UMR 5571, 38041 Grenoble Cedex 09, France\\
  \email{nadege.meunier@obs.ujf-grenoble.fr}
             }

\offprints{N. Meunier}

   \date{Received 11 February 2009 ; Accepted 27 April 2009}

\abstract{The correlation between Ca and $H\alpha$ chromospheric emission, known to be positive in the solar case, has been found to vary between -1 and 1 for other stars. }
{Our objective is to understand the factors influencing this correlation in the solar case, and then to extrapolate our interpretation to other stars. }
{We characterize the correlation between both types of emission in the solar case for different time scales. Then we determine the filling factors due to plages and filaments, and reconstruct the Ca and $H\alpha$ emission to test different physical conditions in terms of plage and filament contrasts. }
{We have been able to precisely determine the correlation in the solar case as a function of the cycle phase. We interpret the results as reflecting the balance between the emission in plages and the absorption in filaments. We found that correlations close to zero or slightly negative can be obtained when considering the same spatio-temporal distribution of plages and filaments than on the sun but with greater contrast. However, with that assumption, correlations close to -1 cannot be obtained for example. Stars with a very low H$\alpha$ contrast in plages and filaments well correlated with plages could produce a correlation close to -1. }
{This study opens new ways to study stellar activity, and provides a new diagnosis that will ultimately help to understand the magnetic configuration of stars other than the sun. }


\maketitle

\section{Introduction}

For the sun, the correlation between the emission in the Ca K and H lines and the emission in H$\alpha$ is wellknown, as both follow the solar cycle \cite[][]{liv07}. However, this may not be the case for other stars, as shown by \cite{cin07}, who found that there was a large dispersion of such correlations. They have studied 109 stars, with a number of observations for each star ranging from a few to a maximum of 22 and covering about 7 years (total span of their observations). They found that for some stars, the correlation could be close to 0 and even negative. They do not provide any interpretation of this result. 

There is therefore a need to understand the nature of this correlation. This type of observation can indeed be used as a diagnosis of stellar activity, but up to now no study has attempted to describe what kind of diagnosis could be derived. What can this correlation teach us about the nature of the stellar activity or the type of dynamo, for example ? 

To understand this, in this first paper (future work will be devoted to more accurate determination of this correlation for a large sample of stars), we study in detail this correlation in the case of the sun. We will first consider its dependence on cycle phase and the influence of the time span used to compute it. The \cite{cin07} observations covered a range smaller than the solar cycle. When considering a short time span, we expect that a smaller correlation will be measured because of the small range of variation of the emission compared to the variability over short time scales. We use the measurements performed by \cite{liv07} at Kitt Peak and covering 19 years to perform this study. This will allow us to determine if such a correlation varies over the solar cycle.  

Then we attempt to understand this correlation by the contribution of emission in plages in Ca and H$\alpha$, which should lead to a correlation very close to 1, and of absorption in filaments, which should decrease the correlation because many of them are not associated directly with active regions and because their contrast is much lower in Ca than in H$\alpha$. We use Meudon spectroheliograms obtained between 1990 and 2002 in both Ca and H$\alpha$ in order to estimate the relative contribution of plages and filaments in the solar case, and then to study their influence on the correlation between the two emissions. This provides clues to study the physics of stellar cycles, as it gives access to more information than the average activity level.  

The outline of this paper is as follows. In Sect.~2, we study in detail 
the correlation between the Ca and H$\alpha$ emission in the solar case, using
Kitt Peak observations. In Sect.~3 we study Meudon spectroheliograms to estimate the contribution of plages and filaments to the observed correlation. These results are used in Sect.~4 to interpret the stellar observations of \cite{cin07}. We conclude in Sect.~5.

\section{Analysis of Kitt Peak observations}

In this section, we consider the sun as a star in order to study the 
correlation between the Ca and H$\alpha$ emissions. For this purpose,
we use the long time series of observations performed at Kitt Peak 
by W. Livingston \cite[][]{liv07}\footnote{These data are available at http://diglib.nso.edu/cycle\_spectra.html}. The sun was observed as a star (full-disk integrated light) using the Fourier Transform Spectrometer at Kitt Peak, thus providing
spectra with a very high spectral resolution (about 500000). The intensity 
in the line core of several lines is also available on the web$^1$.

\subsection{Ca and H$\alpha$ emission variations}

\begin{figure} 
\vspace{0cm}  
\includegraphics{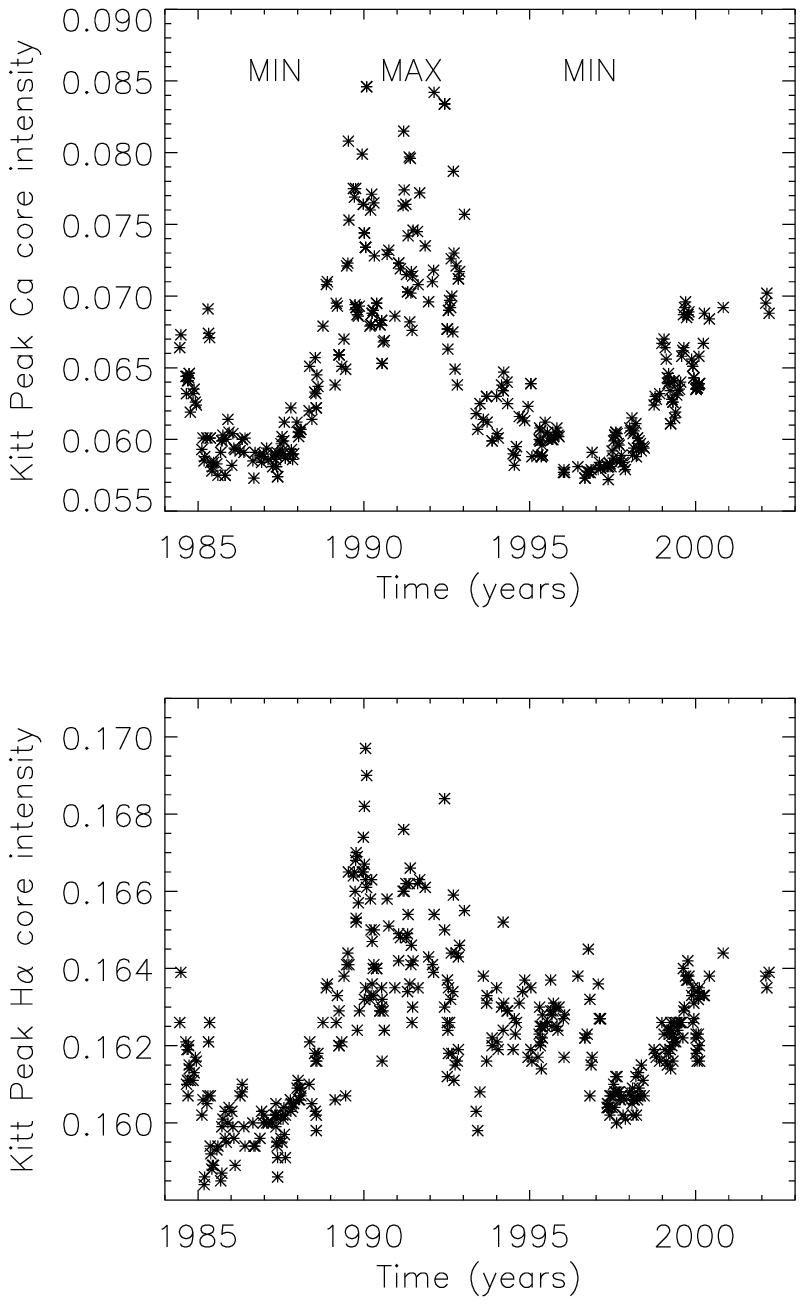}
\caption{{\it Upper panel} : Emission in the Ca line versus time (from W. Livinsgton). {\it Lower panel} : Emission in the H$\alpha$ line versus time (from W. Livingston).  }
\label{varKP}
\end{figure}

We consider the intensity in the core 
for each of these lines, normalized by the continuum intensity 
\cite[see][for more detail]{liv07}. Measurements cover the range 1984--2003 
and consist of 383 observations. They therefore cover 
one and a half cycles, and are evenly spread over the full time range. 
Fig.~\ref{varKP} shows the emission in both lines as a function of
time. These variations are well correlated for example with the activity level
defined by the spot number, as already known \cite[][]{liv07}: the correlation with the daily spot number is 0.86 for Ca and 0.78 for H$\alpha$.

\subsection{Correlation between the Ca and H$\alpha$ emissions}

\begin{figure} 
\vspace{0cm}  
\includegraphics{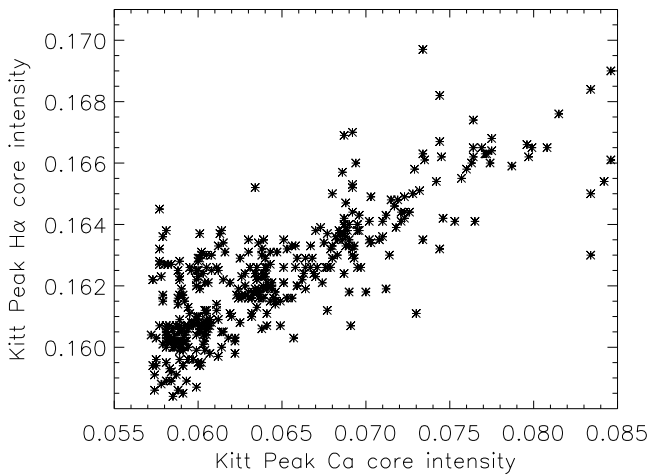}
\caption{Emission in the H$\alpha$ line versus emission in the Ca line (from W. Livingston), from the 1984-2003 complete series. The correlation is 0.806 and the slope 0.27. }
\label{corrKP}
\end{figure}

\begin{figure} 
\vspace{0cm}  
\includegraphics{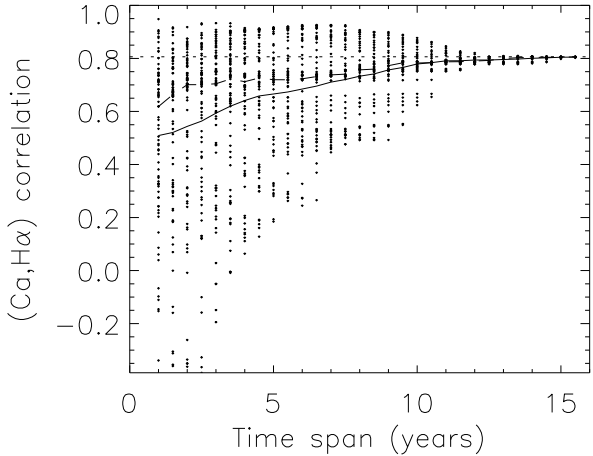}
\caption{Correlation between the emissions in the Ca and H$\alpha$ lines, versus the time span. Each dot corresponds to a different cycle phase. The solid line represents the average and the dashed line the median. The horizontal dotted line corresponds to the correlation computed on the whole data set. }
\label{varcorrKP}
\end{figure}

\begin{figure} 
\vspace{0cm}  
\includegraphics{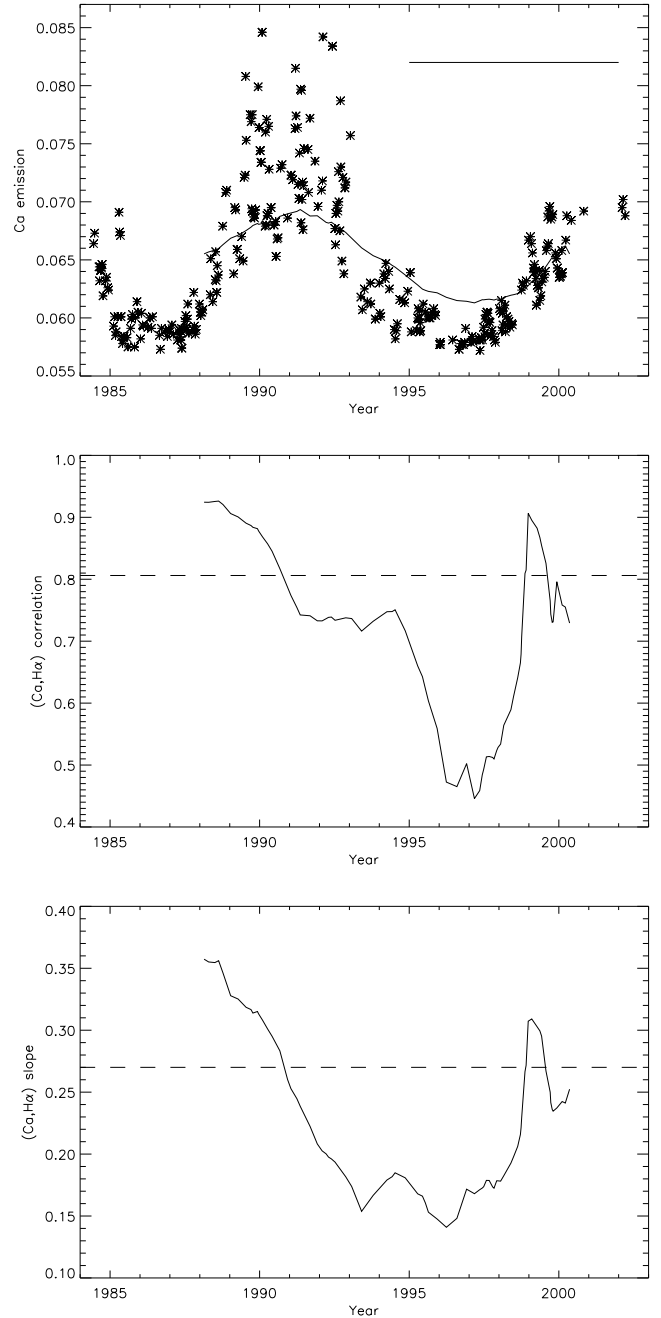}
\caption{{\it Upper panel} : original emission in the Ca line versus time (stars) and average over a time span of 7 years (solid line). The time corresponds to the middle of the 7 year span. {\it Middle panel} : Correlation between the Ca and H$\alpha$ emission versus time for a time span of 7 years. The horizontal dashed line corresponds to the full data set. {\it Lower panel} : Slope of the H$\alpha$ emission versus the Ca emission for the same conditions. }
\label{varKP7}
\end{figure}

We now study in more detail the relation between these emissions. 
We define the correlation factor between the two emissions ${\rm E_{Ca}}$
and ${\rm E_{H\alpha}}$ as~: 
\[ C=\frac{ \sum{ ({\rm E_{Ca}}-\overline{\rm E_{Ca}})({\rm E_{H\alpha}}-\overline{\rm E_{H\alpha}})  }}{   \sqrt{\sum{ ({\rm E_{Ca}}-\overline{\rm E_{Ca}})^2}}  \times \sqrt{\sum{ ({\rm E_{H\alpha}}-\overline{\rm E_{H\alpha}})^2}}  }  \]

We find a correlation of 0.806. The slope of the Ca emission versus the H$\alpha$ emission is 0.27. This means that there is a strong correlation between the two, 
but it is different from 1. Departure from 1 could be due to noise in the
data and/or to physical causes, which will be discussed later.   

In Fig.~\ref{varcorrKP}, we study how the correlation varies with the phase of the solar cycle. We compute the correlation
between the two emissions for different time lags and different starting
times (i.e. different cycle phases) in the series.  For
each time span, all dots correspond to a different starting time in the
series, covering the whole range (minus the time span). For short time spans,
very small correlations can be observed, including correlations close to zero 
or slightly smaller ($\sim$-0.3), i.e. correlations are biased toward values smaller than the one obtained on long time scales.
The correlation for our whole time serie is reached only for time spans close to the cycle length or larger. 

Fig.~\ref{varKP7} shows the variation of the correlation over time, for a
time span of 7 years \cite[as used by][]{cin07}. There is a strong relation between this
correlation and the cycle phase. The correlation is indeed higher than average at 
the end of the ascending phase of the cycle (above 0.9) 
and much lower (about 0.45) at cycle minimum. Furthermore, the curve is 
not symmetric. For a given activity level, the correlation is higher during 
the ascending phase of the solar cycle compared to the descending phase. 
The behavior is similar for the slope.

\section{Meudon spectroheliogram analysis}

\begin{figure*} 
\vspace{0cm}  
\includegraphics{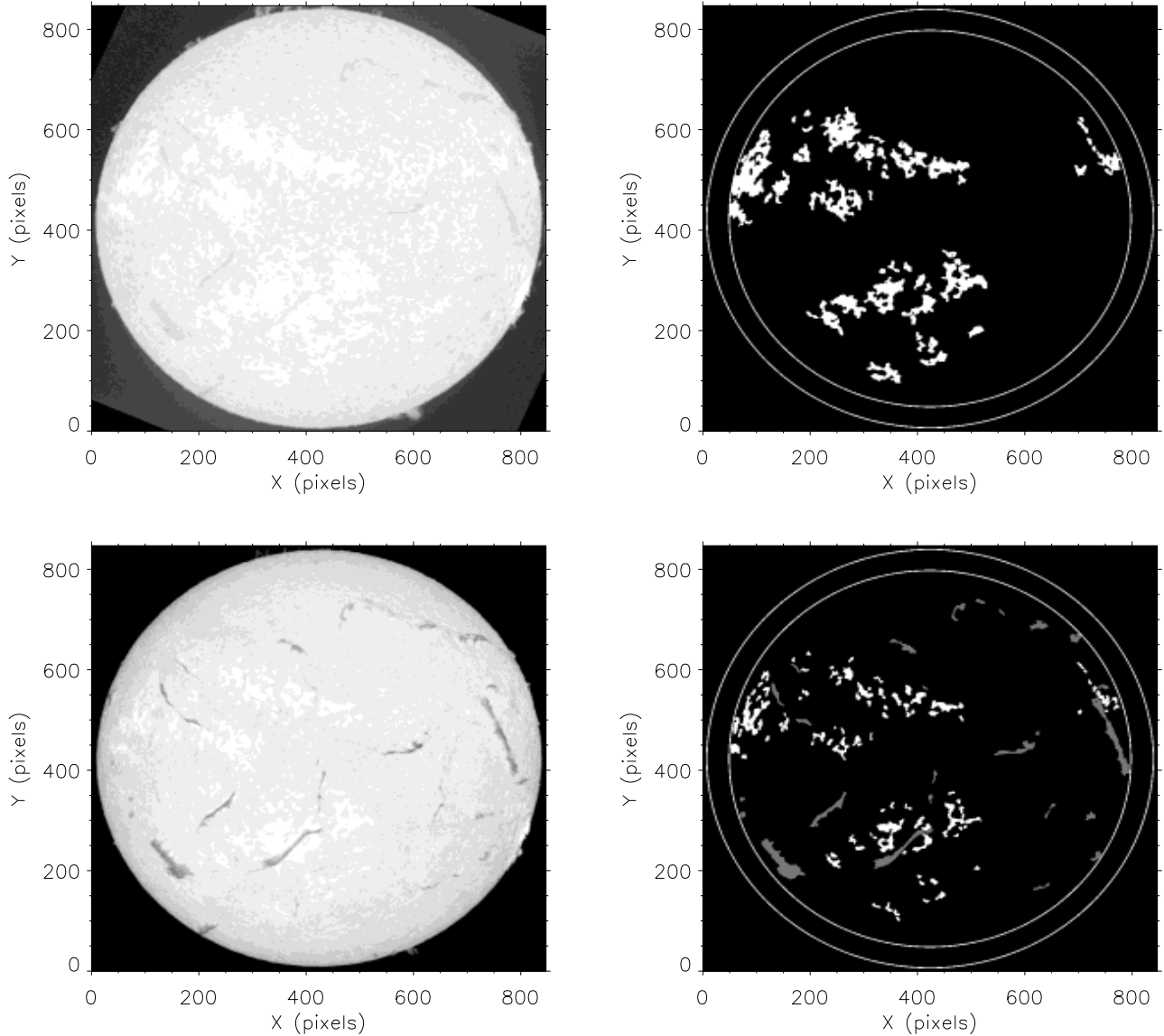}
\caption{{\it Upper panel}: A Ca spectroheliogram (left) and the corresponding segmented image (right). The outer ring shows the solar limb, and the inner ring the limit at 0.9$\times$R$_{\odot}$ used to select structures (plages in that case). {\it Lower panel}: Same for H$\alpha$, with the bright structures corresponding to plages and the grey structures to filaments. }
\label{spectro}
\end{figure*}

\subsection{Contributions to the correlation between the Ca and H$\alpha$ emission}

We now explore possible causes for a correlation of less than 1
between the two emissions in the solar case.
Active regions both show a strong emission in Ca and H$\alpha$ due to plages and their 
position on the solar disk is similar, as shown in Fig.~\ref{spectro}. However, the surface covered by the 
H$\alpha$ plages is slightly smaller than the surface covered in Ca, 
as the latter are formed higher in the chromosphere. The chromospheric network
is also more prominent in Ca than in H$\alpha$. We therefore expect
a small dispersion between the two filling factors, that should lead to a correlation
slightly lower than 1. 

An important difference between the two is the presence of 
dark filaments at the surface, mostly visible in H$\alpha$. They are also  
slightly absorbing in Ca, but with a much lower contrast. When considering the integrated emission, their contribution will therefore be significant for H$\alpha$ but not for Ca : they will decrease the correlation between the Ca and H$\alpha$ emissions towards zero, because they do not coincide exactly with plages. On the other hand, if they were well correlated with plages, a strong filament contrast would lead to an anti-correlation. Filaments are cool and dense gas structures in the corona, maintained against gravity by the presence of aa  magnetic field. They are formed at the boundary between regions of opposite-polarity line-of-sight magnetic fields, either in the quiet sun or in active regions \cite[see][for a review]{martin98}. 

There are two different categories of filaments. 
Some are present in active regions, and these should naturally correlate well with
the activity cycle both on short (days) and long time scales (cycle). However, 
they are usually small and very active, and therefore do not live very long.
Therefore their contribution to the correlation should be small, as these properties are not compensated by very large numbers. Furthermore, 
those that are stable in active regions
represent a small number, and they seem to show no correlation at all with the solar cycle \cite[][]{mackay08}. 

Secondly, there are also many large and stable quiescent filaments outside active regions. These filaments should therefore contribute significantly to the correlation. 
They seem to be correlated with the solar cycle \cite[][]{mouradian94,li07,mackay08}, but again at a large scale (the cycle) and not necessarily at the scale of days or weeks.  Even if their presence is related 
to active regions and in some way ultimately to large regions that are  
predominantly unipolar, the correlation may be low on small time scales,
as there is a time delay between the active regions and the formation of these unipolar network 
regions. 

\subsection{Data analysis}

\begin{figure} 
\vspace{0cm}  
\includegraphics{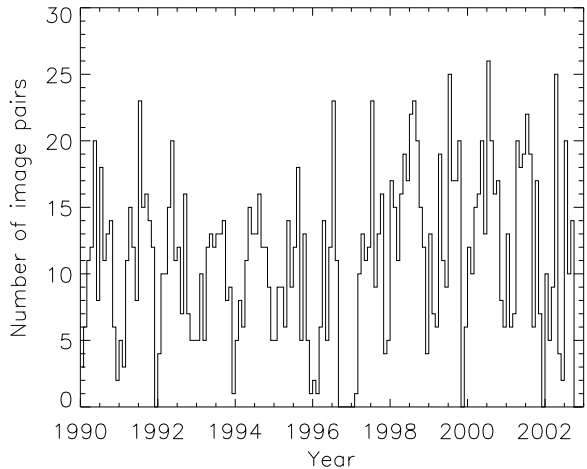}
\caption{Number of image pairs in the Meudon data set}
\label{nbim}
\end{figure}

To estimate the contribution of plages and filaments,
we measure their filling factor
over the solar cycle using Meudon spectroheliograms (provided by BASS2000\footnote{http://bass2000.obspm.fr/home.php}). As mentioned above, the 
filament contrast in Ca is small, and we neglect it in the following. 
We therefore measure the plage filling factor both in Ca (taken at the center of the Ca II K line and showing the upper layer of the chromosphere) and 
in H$\alpha$ images, and the filament filling factor in H$\alpha$ images only. 
We have used 
numerized data obtained between 1990 and 2002, i.e. covering an 11-year solar cycle. 
Fig.~\ref{nbim} shows the number of image pairs (Ca and H$\alpha$) for each 30 day period. There are 1690 pairs in our data set. 
There are naturally more images during the summer,
due to the better weather. There are no large gaps in the data set, except for a few months in 1996, 
and the whole period is therefore well sampled.  

It is beyond the scope of this paper to perform a very sophisticated
region extraction such as those realized by 
\cite{fuller05,ipson05,bernasconi05,aboudarham08} or \cite{scholl08}, as their purpose was to 
identify complete filaments as a whole, with Virtual Observatory
applications in mind (building of synoptic maps for example, in which the 
different pieces constituting a filament must be associated and tracked in time).  
Here we are mainly interested in the total area covered by 
these structures at a given time, therefore there is no need to associate them. This considerably simplifies the segmentation step. 
For the same reason, we do not attempt to use all data, and 
have therefore selected the best images (reasonable seeing, no clouds,
no large dust line). Therefore our algorithm, although presenting strong 
similarities with these previous works, is more simple. It is constituted of the following steps:
center-to-limb darkening correction; analysis of the intensity histograms in order to determine a threshold (ensuring robustness of the final filling factor over time); application of this threshold to the data; application of a thinning procedure; analysis of the segmented data to extract the size of the structures.
Structures smaller than 10 pixels are eliminated (see further for more discussion on the size threshold).

A size selection is performed to eliminate the smallest structures, as we focus on plages and active network to reduce the noise. This size threshold is applied only to the Ca structures (and fixed at 120~Mm$^2$). We have tested thresholds between 30~Mm$^2$ (corresponding to about 10 pixels and therefore very noisy) and 270~Mm$^2$ (corresponding to small active regions). The chosen threshold corresponds to the smallest structures that did not significantly influence the results and allows us to take into account a significant proportion of the magnetic network.  
As the very small structures have a lower contrast \cite[][]{worden98}, it is not a problem to eliminate them at this stage.
Then we only keep structures that are overlapping in both wavelengths. This step makes the procedure robust by eliminating small features that would appear in one of them only due to the threshold choice, if their size were to be close to the threshold.  
For the same reason, we have selected filaments with the same threshold (sizes larger than 120~Mm$^2$), but the influence is not significant either because the smallest features do have a very low contrast as well and therefore do not influence the integrated emission. 
As shown in Fig.~\ref{spectro}, large filaments are well defined, however we do miss the thinnest filaments, which are present mostly in active regions. Because these are much thinner and short-lived, they are the filaments which contribute the less, so it is not a critical point (see Sect.~3.1 for a discussion). 

Finally, a filling factor is computed using the selected structures (surface divided by the total surface of the sun on the image). We are mostly interested in the variability of these filling factors and the correlation between them, more than in the absolute values of the filling factors themselves, so the exact thresholds used to extract the structures is not as sensitive as it would be to get the exact size of these structures for example.

\subsection{Filling factors over the cycle for active regions and filaments}

\begin{figure} 
\vspace{0cm}  
\includegraphics{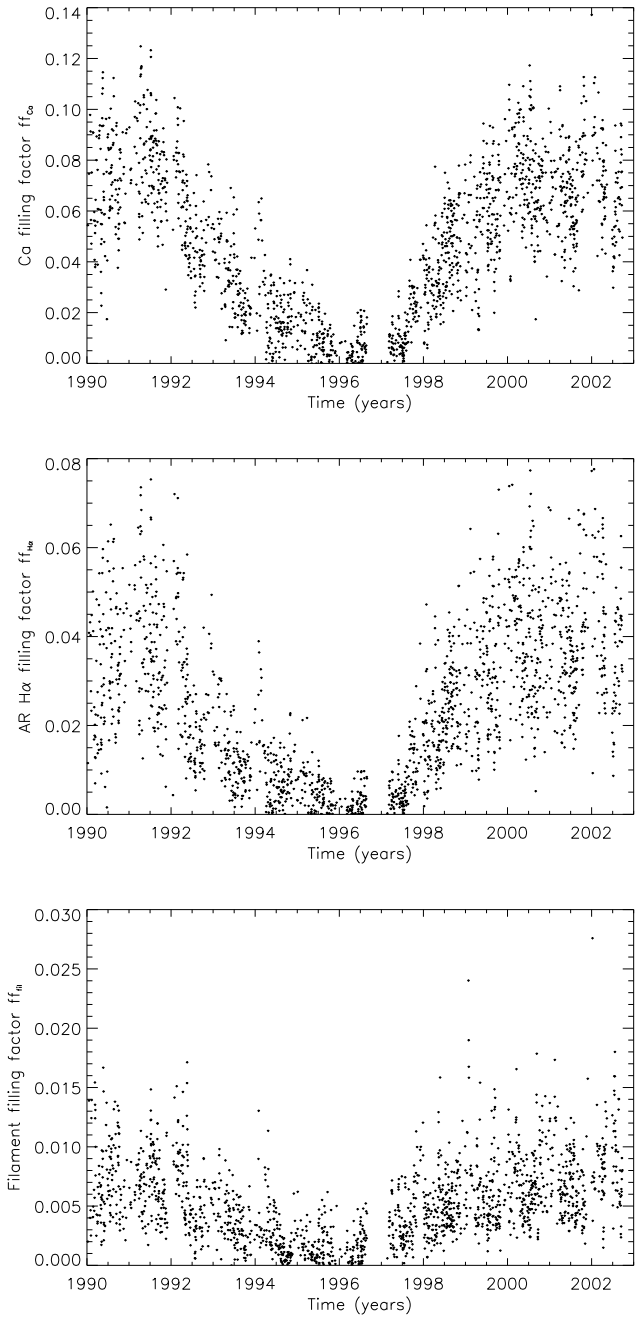}
\caption{ {\it Upper panel} : Filling factor of Ca II K plages. {\it Middle panel} : Filling factor of H$\alpha$ plages. {\it Lower panel}~: Filling factor of filaments.   }
\label{ffmeudon}
\end{figure}

\begin{figure} 
\vspace{0cm}  
\includegraphics{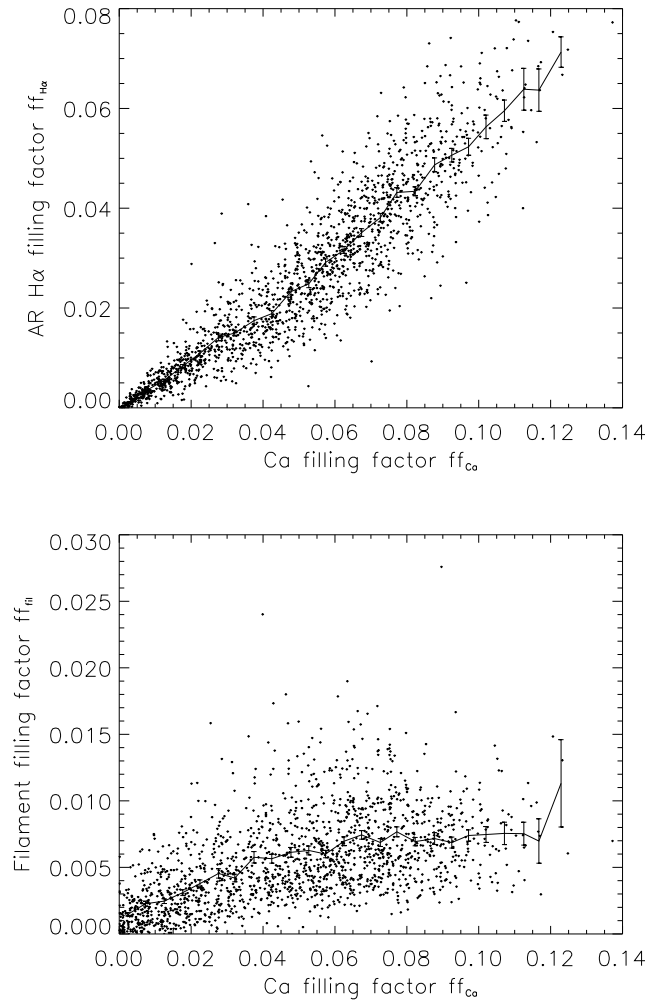}
\caption{ {\it Upper panel} : Filling factor of H$\alpha$ plages versus filling factor of Ca II K plages. {\it Lower panel} : Filling factor of filaments versus filling factor of Ca II K plages.   }
\label{ff2meudon}
\end{figure}

\begin{figure} 
\vspace{0cm}  
\includegraphics{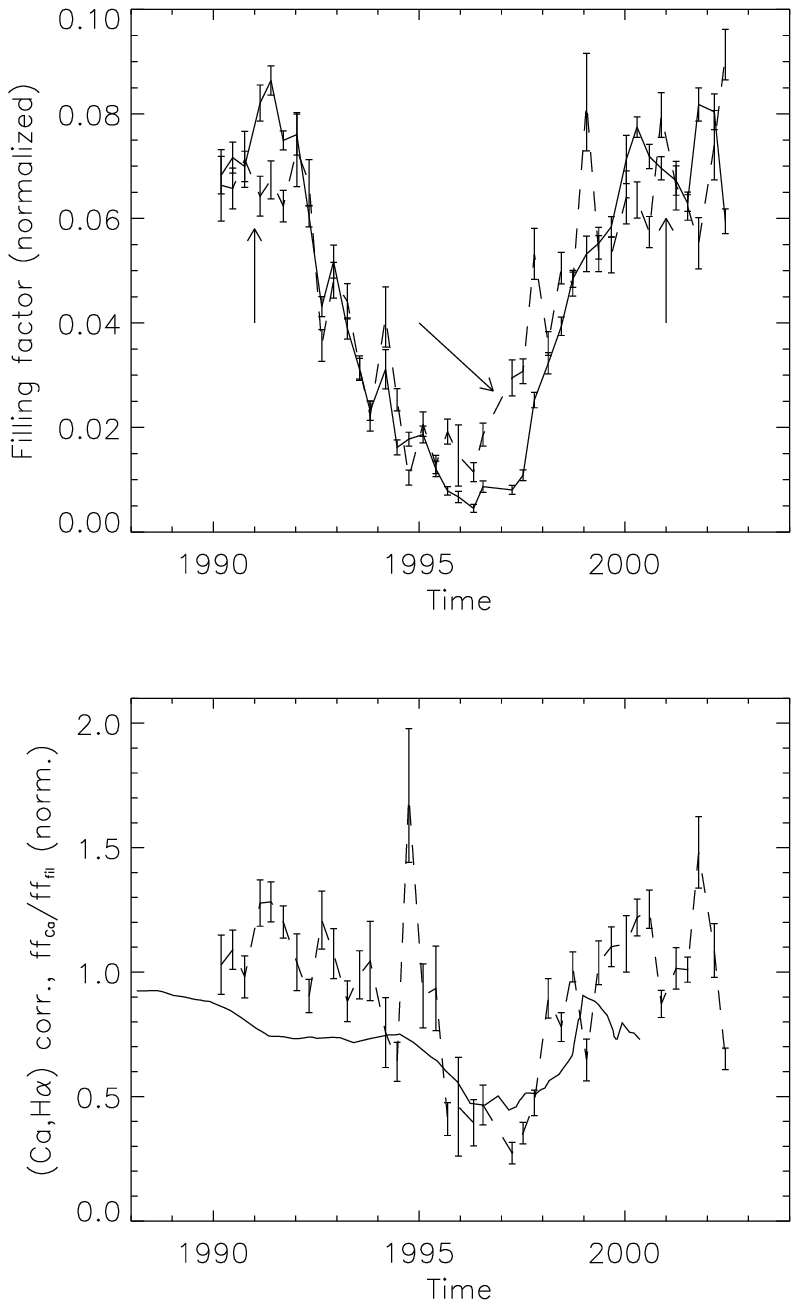}
\caption{ {\it Upper panel} : Ca filling factor ff$_{\rm Ca}$ (solid line) and filament filling factor ff$_{\rm fil}$ normalized to the same median (dashed line), both averaged of 4 rotation periods, versus time.  
{\it Lower panel} : Correlation between the Ca and H$\alpha$ Kitt Peak emission versus time (solid line, from Fig. 4, middle panel) superimposed on the ratio (dashed line) between the two upper curve (ff$_{\rm Ca}$/ff$_{\rm fil}$ with a normalization of ff$_{\rm fil}$ as in the upper panel). }
\label{fil_cycle}
\end{figure}

Fig.~\ref{ffmeudon} shows the resulting filling factor versus time for the 
3 types of structures: plages identified on Ca II K images (ff$_{\rm Ca}$), plages 
identified on H$\alpha$ images (ff$_{\rm H\alpha}$), and filaments identified on H$\alpha$ images 
as well (ff$_{\rm fil}$). 
As expected, we observe a strong correlation of 0.92 between the plage filling factor 
in Ca images and in H$\alpha$ images. The filament filling factor
is also higher at cycle maximum, but the correlation
with the plage filling factor is lower, with values of  
0.52 for Ca and 0.53 for H$\alpha$.

Fig.~\ref{ff2meudon} also shows ff$_{\rm H\alpha}$ and ff$_{\rm fil}$ versus 
ff$_{\rm Ca}$. The correlation is quite clear in both cases. For H$\alpha$
plages, the relation is close to linear with a change in slope
for ff$_{\rm Ca}$ around 0.04: at higher activity levels, ff$_{\rm H\alpha}$ increases faster than ff$_{\rm Ca}$. This may be due to the largest structures at that time, with therefore a ratio closer to one between the size of the regions observed in Ca and H$\alpha$. As for filaments, a linear increase is
observed for ff$_{\rm Ca}$ up to 0.06, and then a saturation (corresponding to cycle maximum): above a 
certain activity level, the filament filling factor  remains constant, which is a new result. This and 
the large dispersion explain the correlation close to 0.5, i.e. far from 1.  
We also note that at cycle minimum, when ff$_{\rm Ca}$ and ff$_{\rm H\alpha}$ both go very close to zero, ff$_{\rm fil}$ is significantly above zero, showing the presence of a significant number of filaments, as already observed \cite[][]{makarov92,mouradian94} . 

The different time scales will be important when interpreting stellar data. On a long time scale (cycle period for example), the filament filling factor is roughly correlated with the activity level (determined from the Ca filling factor for example), as already pointed out by \cite{mackay08} for a few selected periods over the cycle. In addition to the saturation at high activity levels found above, the correlation is not close to 1 due to a large dispersion, and this means that at small time scales (rotation period) the correlation could be much closer to zero. Finally, Fig.~\ref{fil_cycle} shows the filling factors for Ca and filaments for an averaging over temporal windows of about 4 rotation periods. On this plot, ff$_{\rm fil}$ has been normalized to ff$_{\rm Ca}$ (same median) to make the comparison easier. The correlation at this large temporal scale is now much closer to 1 (0.87) due to the temporal smoothing. The value of 0.92 (as for ff$_{\rm Ca}$ and ff$_{\rm H\alpha}$ obtained above) is found for a temporal smoothing of about 200 days.  There are, however, a few noticeable differences. At cycle minimum and during the beginning of the ascending phase, the surface covered by filaments increases faster than the plage filling factor. On the other hand, at cycle maximum, the peaks are not as sharp, especially for the 1990--1991 maximum, which is related to the saturation above. The cycle maximum in 2000--2002 is known to exhibit two peaks, as shown by the ff$_{\rm Ca}$ : this is not observed for filaments, with in fact a peak in between, shifted by about 200--300 days from the first peak. 
Fig.~\ref{fil_cycle} also shows that the dip in the ratio between the plage and filament filling factors observed around 1996--1997 corresponds to the low correlation factor observed in the Kitt Peak emission at that time, which reinforces our interpretation of the correlation evolution by the presence of filaments.

%\begin{minipage}
%\renewcommand{\footnoterule}{}
\begin{table}
\begin{center}
\caption{Correlations and parameters from a fit in the Kitt Peak data or from the literature.   }
\label{tabres}
\begin{tabular}{ccccc}
\hline \hline
Type &  Variable &  Kitt Peak & Fit &  Literature \\ \hline
Obs. &  Corr(ff$_{\rm Ca}$,ff$_{\rm H\alpha}$) & ... & 0.923  & 0.923  \\
&  Corr(ff$_{\rm Ca}$,ff$_{\rm fil}$) &  ... & 0.524 & 0.524   \\
&  Corr(ff$_{\rm H\alpha}$,ff$_{\rm fil}$) & ... & 0.520  & 0.520   \\ \hline
Param. &  C$_{\rm Ca}$ & ... & 0.20  &  0.70  \\ 
&  C$_{\rm H\alpha}$ & ... & 0.13  & 0.35  \\
&  C$_{\rm fil}$ & ... &  0.30 &     0.20  \\ \hline
Constr. &  Corr(E$_{\rm Ca}$,E$_{\rm H\alpha}$) & 0.805  &  0.805 & 0.913 \\
&  Slope(E$_{\rm Ca}$,E$_{\rm H\alpha}$) & 0.267  &  0.268 &  0.260 \\
&  Ampl. E$_{\rm Ca}$ & 0.028 & 0.028  &  0.096   \\
&  Ampl. E$_{\rm H\alpha}$ & 0.012 & 0.013  &  0.027  \\
&  $\chi^2$ & ... & 0.0009 &  0.109 \\  \hline
\end{tabular}
\end{center}
\note{The first 3 lines provide the correlation between the observed filling factors. The following set of lines gives the contrasts which have been used (either from a fit on from the literature). The last set of lines shows the constrains that can be used to either constrain the fit or, in the last column, to test the contrasts found in the literature, compared to the values expected from the Kitt Peak observations. }
\end{table}
%\end{minipage}

\subsection{Reconstruction of the emission}

\begin{figure} 
\vspace{0cm}  
\includegraphics{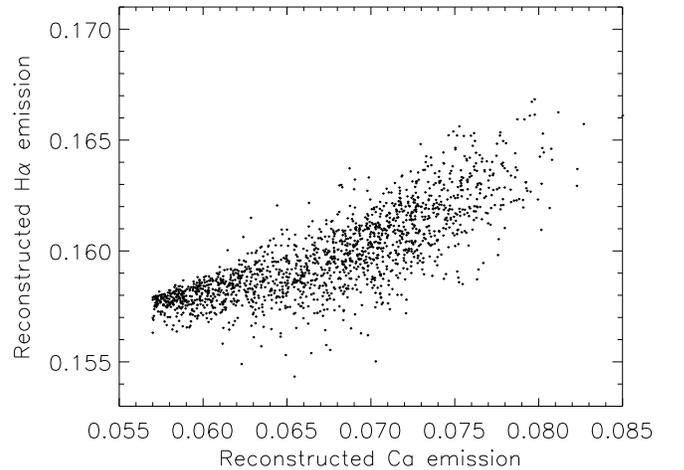}
\caption{Reconstructed emission in Ca II K versus the reconstructed emission in  H$\alpha$.}
\label{reconst_corr}
\end{figure}

In using contrasts for plages in Ca and H$\alpha$, and for filaments 
in H$\alpha$ (C$_{\rm Ca}$, C$_{\rm H\alpha}$ and C$_{\rm fil}$ respectively), it is possible to reconstruct the emissions in Ca and H$\alpha$ as defined by :
\[ {\rm E_{Ca}= C_{Ca} \times ff_{Ca} } \] 
\[ {\rm E_{H\alpha}= C_{H\alpha} \times ff_{H\alpha} - C_{fil} \times ff_{fil}}. \]
In the following, we have assumed constant contrast over time. We have also considered the same contrast for structures of all size. We know that for bright structures, there is a variability depending on their size from plages down to the network \cite[][]{worden98}, however in this work we focus mostly on plages (see Fig.~\ref{spectro}), so we do not expect a large effect.  

Unfortunately, the available Meudon data alone do not allow us to compute the plage and filament contrast, as there is no intensity calibration. 
It is, however, possible to determine them using both Meudon and Kitt Peak data. 
We have estimated the best 
contrasts C$_{\rm Ca}$, C$_{\rm H\alpha}$ and C$_{\rm fil}$ that would
lead to reconstructed emissions varying over a range similar to that observed at Kitt Peak. We also constrain these contrasts by adjusting them in order to 
obtain a correlation and a slope between the reconstructed emissions as close as possible to the Kitt Peak values. 
We therefore minimize 
a $\chi^2$ which takes into account these three factors, equally weighted. 
The smallest $\chi^2$ then provides the best estimates for the three contrasts. 
Using these contrasts and the observed filling factors, we derive 
the reconstructed emissions ${\rm E_{Ca}}$ and ${\rm E_{H\alpha}}$.
We find contrasts that 
reproduce the Kitt Peak observations (see Table~\ref{tabres}). 
These contrasts correspond to a small band pass, as we are using Kitt Peak intensities in the line core. 
Fig.~\ref{reconst_corr} shows E$_{\rm H\alpha}$ versus E$_{\rm Ca}$ in that case and can be compared 
to the Kitt Peak observations in Fig.~\ref{corrKP}. We note that despite
the same amplitude, correlation and slope, the behavior is slightly different, as Kitt Peak observations exhibit the largest dispersion for small activity levels, while here the largest dispersion is for large activity levels.

Another possibility is to use the contrasts published in the literature. 
Unfortunately, we did not find any large and systematic studies that would provide all necessary 
contrasts. One reason is that they depend significantly  on the band pass
and that 
there are few studies on a large sample. For Ca structures, we have used the value of 0.7 
deduced from \cite{worden98}, who obtained a variation with the type of structure
(from the network to plages). \cite{ortiz05} also observed a variation of the Ca 
contrast as a function of the line-of-sight magnetic field. We considered the value of 0.35 
for  C$_{\rm H\alpha}$, assuming a factor of two between Ca and $H\alpha$ \cite[][]{kono75}. This ratio of 2 is close to what we obtained above with the fit. The filament contrasts show a very large dispersion in the literature and we have taken an average value of 0.2 deduced from \cite{maltby76} and \cite{chae06}. As shown  in Table~\ref{tabres}, the $\chi^2$ is much worse in that case, and it is not possible to reproduce the observed amplitude of variation and the correlation, as the plage contrasts are too high and the filament contrast too low. 
This may be due to the difference between the band pass at which they were measured \cite[0.5~\AA$\;$ for example in the case of][]{worden98} and the Kitt Peak observations we are using. 

\section{Toward diagnosis of stellar activity}

\subsection{Distribution of stellar and solar correlations}

\begin{figure} 
\vspace{0cm}  
\includegraphics{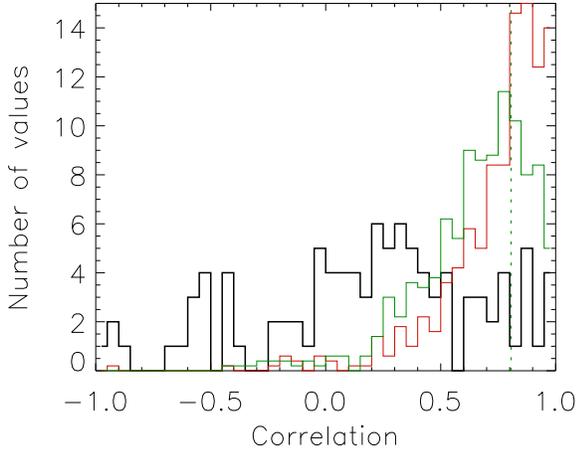}
\caption{Comparison of the distribution of correlations obtained by Cincunegui et al. (2007) for a sample of 109 stars (solid line) and for the sun viewed from the equator normalized to the same number of stars (this work) for various phase of the solar cycle and a similar time span, i.e. 7 years (red line).  The green line is derived from the same conditions as for the red line, but takes into account a uniform distribution of inclination angles between 0 and 90$^{\circ}$ for a latitudinal extent of [-30$^{\circ}$,30$^{\circ}$] for plages and [-60$^{\circ}$,60$^{\circ}$] for filaments (see Sect.~4.1 for details).  } 
\label{compdistcorr}
\end{figure}

\begin{table}
\begin{center}
\caption{Filament contrasts for which the correlation is equal to zero.}
\label{tabres2}
\begin{tabular}{ccc}
\hline \hline
 Contrast & Fit &  Literature \\ \hline
 C$_{\rm Ca}$ &  0.20  &  0.70  \\ 
 C$_{\rm H\alpha}$  & 0.13  & 0.35  \\
 C$_{\rm fil}$ &   0.30 &     0.20  \\ \hline
 C$_{\rm fil}$(corr=0)  & 1.16  &  3.08   \\
 C$_{\rm fil}$(corr=0)/C$_{\rm fil}$   &  3.9 & 15.4  \\  \hline
\end{tabular}
\end{center}
\note{Given the fitted contrasts for the three types of structures or those found in the literature, a filament contrast for which the correlation is equal to zero is given, in absolute value and relative to the solar value (see Sect. 4 for details). }
\end{table}

Fig.~\ref{compdistcorr} shows the distribution of correlations obtained by 
\cite{cin07}. They observe a significant number of stars with correlations 
close to zero or negative. 
Superimposed on it, we show the distribution in the case of the sun, 
for a time span of 7 years (red line), and covering different phases of the solar cycle, 
deduced from the dots of Fig.~\ref{varcorrKP} for that time span. Several conclusions can be 
drawn from this comparison. 

First, the two distributions are very 
different, as the tail toward small correlations is of small amplitude 
in the case of the sun, compared to the number of cases around 0.8. 
However, \cite{cin07} provide no uncertainties on the correlations, and some of these are large (given the available plots). However, large uncertainties probably do
not suffice to explain the difference in distribution. This means that the
activity of the sun is probably not typical of that of other stars, possibly due to filaments. A possible source of discrepancy could be the heterogeneity in the sample studied by \cite{cin07} in terms of stellar type. 

It is also necessary to take into account the large range of inclination angles present in stellar observations. In the solar case, the latitudinal extent of plages and filaments is different, so one can expect a different contribution from these two types of structures depending on the inclination angle. If considering that plages extend up to 30$^{\circ}$ and filaments up to 60$^{\circ}$, for example, and assuming that the filling factor at each time step is attributed to that latitudinal range, a correlation of 0.8 (sun viewed from the equator) would fall to 0.75 for an inclination angle of 45$^{\circ}$ and to 0.49 for a pole-on orientation. For a uniform distribution of inclination angles between 0 and 90$^{\circ}$, the distribution shown in Fig.~\ref{compdistcorr} (red line) should then be smeared toward lower correlations and is shown in green.  

The second conclusion is
that despite these differences, the solar distribution exhibits small correlations in a few cases. This means that if stellar observations do not cover a full activity cycle (or more generally, the full range of activity existing for that star), we are likely to significantly underestimate the correlation with respect to the one that would be obtained when considering a full cycle. In that case, the only way to consider the data is a statistical analysis of a large number of stars, i.e. 
via the distribution of correlations. Indeed, for a given star, the correlation
may be low (for example close to zero) even if in reality the star has a correlation computed over the whole cycle similar to the solar one. 
If the cycle length is known, as well as the phase, it may be possible, in principle, to apply a correction deduced from the plot of 
Fig.~\ref{varKP7} (or a similar one if the time span of the observations
is different). In that case, a distribution of corrected correlations could
be produced, that would give an indication of the true correlations over the
whole activity range. However, such corrections would assume a behavior similar to the sun (in particular, relative contrast).  

\subsection{Influence of the filament contrast on the correlation}

\begin{figure} 
\vspace{0cm}  
\includegraphics{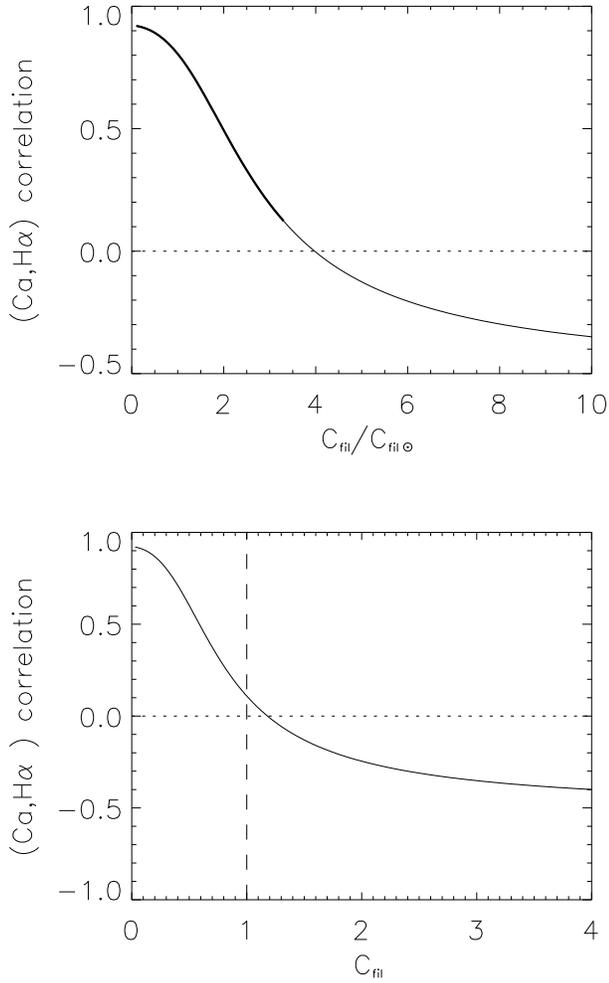}
\caption{{\it Upper panel}: Correlation between the reconstructed emission in Ca II K and H$\alpha$ for various filament contrasts (relative to the fitted solar value). {\it Lower panel}: Same versus the actual filament contrast, without the normalization to the solar value.}
\label{corr2}
\end{figure}

We now consider cases for which the correlation is computed over the whole cycle. 
It would also be interesting to know what the sensitivity is of 
the correlation for filament contrasts differing from the solar contrasts 
found above. Fig.~\ref{corr2} shows the correlation between the reconstructed emission in Ca and H$\alpha$ for various filament contrasts, while keeping the same contrasts for plages in Ca and H$\alpha$. For low contrast, the correlation is naturally close to 1. For increasing filament contrast, the correlation decreases, crosses zero for a certain contrast, and then saturates for values around -0.4.  
Table~\ref{tabres2} shows the contrast necessary to reach a correlation of zero as well as the ratio to
the solar contrast.  A contrast 3.9 times the solar 
value of 0.30 (i.e. 1.16) and the same distribution in time and position of the 
filaments would lead to a correlation close to 0. The factor is, of course, the 
same when applied to
the area covered by the filaments instead of their contrast. 
When using the contrast obtained from the literature, the filament contrast necessary to reach a zero correlation is much higher. 
A contrast of 1.16 is larger than 1, which means that a correlation
of zero for the same spatio-temporal organization of the filaments 
can be reached only if the filling factor of filaments is increased. 
We see that, for example, for a factor 2 increase in contrast and in size,
which is not very extreme (the observed contrast of individual solar filaments 
varies by a factor of 3-4), at least such a zero correlation is reached. 

However, we cannot obtain a strong negative correlation below $\sim$-0.4, 
and such values correspond to extreme conditions compared to the sun. 
Therefore, the spatio-temporal organization of filaments must be 
significantly different if the presence of filaments explains 
the Ca - H$\alpha$ relation observed in other stars. 
For example, filaments that would be more strongly in phase with plages
and with a contrast dominating that of H$\alpha$ plages could lead 
to an anti-correlation.

\begin{figure*} 
\vspace{0cm}  
\includegraphics{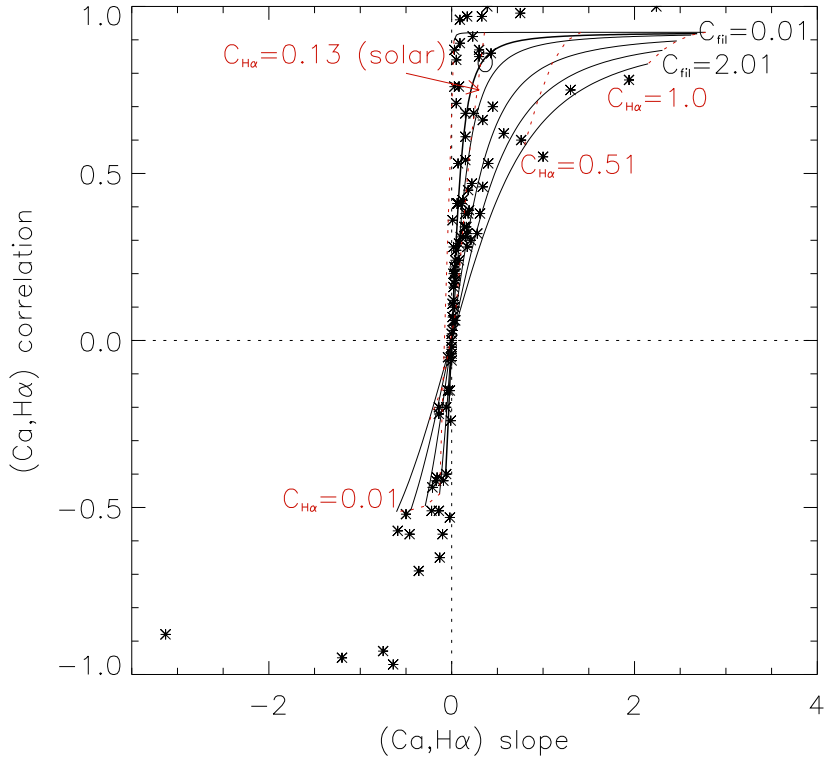}
\caption{Correlation between the Ca and H$\alpha$ emission versus the slope, for the 
Cincunegui et al. (2007) results (stars) and the sun (circle). A model corresponding to the solar C$_{\rm fil}$ and a variable C$_{\rm H\alpha}$ is superimposed (thick line), as well as 5 models (thin lines), each of them with a different C$_{\rm fil}$ (from a low filament contrast for the upper curve to a high filament contrast for the lower curve). For each curve, C$_{\rm H\alpha}$ varies from 0.01 to 1. C$_{\rm Ca}$ is constant (fitted solar value of 0.20).  }
\label{m_rho}
\end{figure*}

\subsection{Influence of the filament and plage contrasts on the correlation}

We now look at the influence of all contrasts on the correlation and the slope
simultaneously. 
Fig.~\ref{m_rho} shows the correlation versus the slopes 
as observed by \cite{cin07} superimposed with the expected correlations and slopes for solar filling factors multiplied by various values of the H$\alpha$ plage contrast (between 0 and 1) and of the filament (between 0 and 2, i.e. allowing a larger filling factor for the filaments). Most stars could then be explained by this interpretation, i.e. with the assumption that plages and filaments have the same spatio-temporal distribution than on the sun but different contrasts. The most notable exceptions are a few stars with correlations very close to -1. There are also a few stars with correlation between 0.92 (the solar case with no filament) and 1 but the uncertainties on the correlation in these stellar observations are larger than this difference. 

This graph showing the correlation versus the slope allows us to explore different physical conditions that may be found on stars of different types (e.g. different temperature and density profiles in the chromosphere). 
The value of the correlation depends significantly on several factors. First, it depends on the ratio between the plage contrast and the filament contrast, as shown in Sect.~4.2. It also varies with the ratio between the plage contrast in Ca and H$\alpha$. These lines are not formed at the same level in the chromosphere, and we could expect stars of different types to exhibit a ratio different from the solar value. For stars with a H$\alpha$ contrast much smaller than the Ca one, the correlation falls rapidly to zero or to negative values.

\section{Conclusion}

The first result of this work  is a precise characterization of the correlation between the 
Ca and H$\alpha$ emission in the solar case : the value of the correlation, dependence on the time scale and cycle phase, and finally interpretation using filaments, which decrease the correlation due to plages. Also, we have found that the filament filling factor is correlated to the activity level (0.5) but with a large dispersion (due to the low correlation at short time scales) and a saturation effect at high activity levels, which is new. 

In addition, we can use these results to estimate what would be expected on stars with a similar spatio-temporal distribution of filaments and plages but different sizes or contrasts. We can explain most of the results obtained by \cite{cin07}, but not all of them. This may reflect physical conditions different from solar, either in chromospheric properties or in the activity cycle. 
However, the uncertainties in the results obtained by \cite{cin07} seem to be quite large and they could be responsible for the observed discrepancy with the solar case. There is therefore a need to study in more detail the precise relationship between the Ca and H$\alpha$ emissions in a large star sample. This is probably at least partially due to the poor temporal sampling. This is also necessary because their sample is quite heterogeneous in stellar types. This is a work in progress.    

If we want to extrapolate to stellar cases with different magnetic configurations and consequently a different spatio-temporal distribution of filaments, we need to obtain a better estimate of the filament behavior over the solar cycle.
We therefore plan to study this using the very large data set of MDI/SOHO magnetograms \cite[][]{Smdi95}. This will allow us to establish a link to the magnetic configuration. 

\begin{acknowledgements}
The Kitt Peak observations kindly have been made available on the web by W. Livingston. We thank W. Livingston and R. Toussaint for their comments on these data. The Meudon spectroheliograms were provided by BASS2000. We thank J. Aboudarham for informations on these data. 
The monthly sunspot number was provided by 
the SIDC-team, World Data Center for the Sunspot Index, Royal Observatory of Belgium, Monthly Report on the International Sunspot Number, online catalogue of the sunspot index: http://www.sidc.be/sunspot-data/, 1984-2003. 
We thank the anonymous referee for his/her comments. 
\end{acknowledgements}

\bibliographystyle{aa}
\bibliography{biblio11823}

\end{document}